# State Drift and Gait Plan in Feedback Linearization Control of a Tilt Vehicle

Zhe Shen and Takeshi Tsuchiya

Department of Aeronautics and Astronautics,
The University of Tokyo, Tokyo, Japan

## Abstract

*To stabilize a conventional quadrotor, simplified equivalent vehicles (e.g., autonomous car) are developed to test the designed controller. Based on that, various controllers based on feedback linearization have been developed. With the recently developed concept of tilt-rotor, there lacks the simplified/equivalent model, however. Indeed, the tilt structure is relatively unusual in vehicles. In this research, we put forward a unique fictional vehicle with tilt structure, which is to help evaluate the property of the tilt-structure-aimed controllers. One phenomenon (state drift) in controlling an over-actuated tilt structure by feedback linearization is presented subsequently. State drift can be easily neglected and is not paid attention to in the current researches in tilt-rotor controllers' design so far. We report this phenomenon and provide a potential approach to avoid this behavior.*

## 1. Introduction

Various controllers are analyzed to control the conventional quadrotors. Some of them are linear. [5] compares two linear controllers in stabilizing the attitude of a quadrotor model. Others are nonlinear. [6, 7] apply geometric controller, which delicately defines the attitude error and provides the stability proof. [8] puts forward a sliding mode controller; it guarantees the stability using Lyapunov criteria. Backstepping controller with cascade structure can be found in [9].

Both linear and nonlinear controllers have their advantages and disadvantages. These pros and cons include the controlling time, stability region, dynamic bias, robustness, calculating cost, etc. Different from solely relying on linear or nonlinear controllers, feedback linearization [2,3] transfers a nonlinear system to a linear system. The linear controller can be subsequently applied to the generated linear system.

Unfortunately, the quadrotor is an under-actuated system where the number of inputs is less than the degrees of freedom. This nature prohibits us from building a full-state controller; the number of the states to be controlled directly is 4 at most, equal to the number of the inputs. Several different selections on these states can be found in [2, 10, 11]. Another problem hindering us from utilizing feedback linearization is the invertibility of the decouple matrix [2, 12]. Some selections of the output combination introduce the singular decouple matrix. [12] details this problem and provides the potential approach to avoid hitting the singular zone.

The recent popular novel quadrotor (tilt-rotor) augments the number of inputs to 8 [4, 13, 14]; the quadrotor not only changes the magnitude of each thrust but also the direction. With this designation, the system becomes over-actuated. Subsequently, several researches focus on controlling this over-actuated system by feedback linearization [4], [15]-[17].

Controlling an over-actuated system by feedback linearization provides the possibility of full-state control. The input, however, might change even if the state has stabilized. This phenomenon is not paid attention to yet.

In this paper, we address the feedback-linearization-based controllers in a simplified tilt vehicle. Controlling a simplified vehicle is a common approach before applying this control method to the complicated system (e.g., UAV) [1]. We present the result with the state drift phenomenon. To avoid this unwanted drift, we decrease the number of the real inputs by gait plan, which is a gait schedule technique in quadruped robots [18].

This paper is structured as follows: Section 2 introduces the dynamics of the fictional vehicle. The feedback-linearization-based controller is designed in Section 3. Section 4 simulates the controller designed and presents the state drift phenomenon. Section 5 proposes gait plan and puts it into the controller design. Section 6 concludes and makes the discussion of the result.

## 2. DYNAMICS OF THE TILT VEHICLE

In this section, we introduce a two-dimensional fictional tilt vehicle (Figure 1). The motion of the vehicle is restricted to a level plane (two-dimensional movement). Two propellers are fixed on the vehicle with the angle $2\theta$ between. The value of $2\theta$ is $\frac{\pi}{3}$.

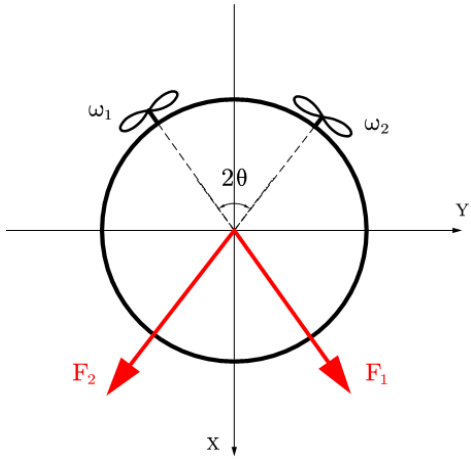 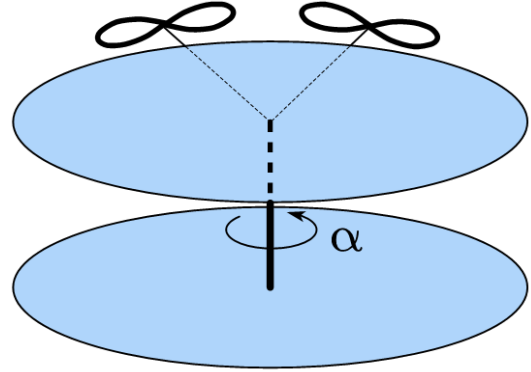

Figure. 1. Two-dimensional Tilt Vehicle     Figure. 2. Tilt Structure

Each propeller can generate the thrust by rotating with angular velocity, $\omega_1$ and $\omega_2$. The trust generated follows Equation (1).

$$F_i = K_{F_i} \cdot \omega_i^2, \ i=1,2 \ (\omega_i \geqslant 0) \tag{1}$$

where $K_{F_i}$ is the thrust coefficient. Its value is $K_{F_i} = 0.001$. This proportional relationship in adopted in the dynamics of UAV control [1-3]. The thrust is always nonnegative. Also, the angular velocity is always nonnegative.

What makes this vehicle special is the tilt structure, as plotted in Figure 2. The propellers mentioned are fixed on the top disc. While the bottom disc contacts the $x-y$ plane directly. The frictional force from the $x-y$ plane is ignored. This structure provides the possibility of changing the directions of the thrusts by tilting the top disc (Figure 3—4) with an angle, $\alpha$.

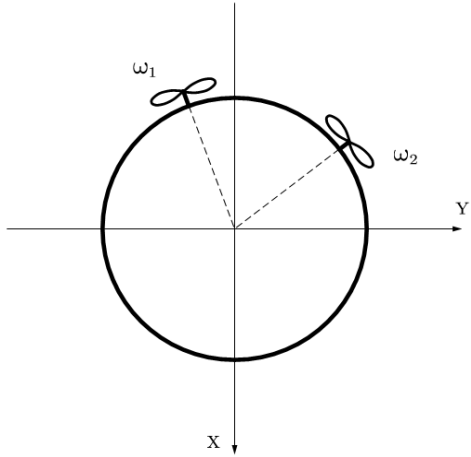
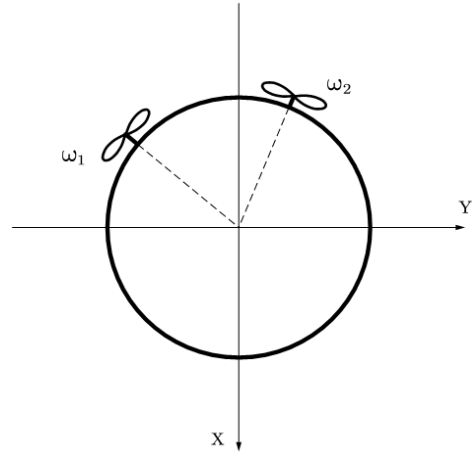

Fig. 3. Tilt to right example        Fig. 4. Tilt to left example

Based on Conservation of Angular Momentum, the relationship between the yaw, $\psi$, and the tilt angle, $\alpha$, is written in Equation (2).

$$\psi = -\frac{I_T}{I_B} \cdot \alpha \tag{2}$$

where $I_T$ and $I_B$ are the rotational inertias of the top disc and bottom disc, respectively. Their values are $I_T = 0.001$, $I_B = 0.002$.

Based on the Newtown's law, we write the position in Equation (3).

$$\begin{bmatrix} \ddot{x} \\ \ddot{y} \end{bmatrix} = \frac{1}{m} \cdot J_{\psi+\alpha} \cdot J_\theta \cdot \begin{bmatrix} \omega_1^2 \\ \omega_2^2 \end{bmatrix} \tag{3}$$

where $m$ is the mass of the entire vehicle. $J_{\psi+\alpha}$ and $J_\theta$ are defined in Equation (4)—(5).

$$J_{\psi+\alpha} = \begin{bmatrix} \cos(\psi+\alpha) & -\sin(\psi+\alpha) \\ \sin(\psi+\alpha) & \cos(\psi+\alpha) \end{bmatrix} \tag{4}$$

$$J_\theta = \begin{bmatrix} \cos(\theta) & 0 \\ 0 & \sin(\theta) \end{bmatrix} \cdot \begin{bmatrix} K_{F_1} & K_{F_2} \\ K_{F_1} & -K_{F_2} \end{bmatrix} \tag{5}$$

It is worth mentioning that both $J_{\psi+\alpha}$ and $J_\theta$ have full rank. Thus, $J_{\psi+\alpha} \cdot J_\theta$ is invertible.

$$|J_{\psi+\alpha} \cdot J_\theta| \neq 0 \tag{6}$$

So far, we have derived the dynamics of this vehicle in Equation (2)—(3). This is a MIMO system. There are three inputs, $\omega_1$, $\omega_2$, and $\alpha$. While there are two outputs, $x$, and $y$. So, it is an over-actuated system.

## 3. CONTROLLER DESIGN (FEEDBACK LINEARIZATION)

Remarks to this MIMO system: Since it is an over-actuated system, we may apply feedback linearization for this system. The other idea is to discard some inputs to let the number of the outputs equal to the number of the inputs. Although Feedback Linearization is widely used the tilt-rotor control, one potential adverse effect (state drift) is always neglected by researchers.

This section develops the feedback linearization controllers based on this over-actuated system model. The result and the state drift phenomenon are given in the next section.

### 3.1. Third Derivative Feedback Linearization

Since the input $\alpha$ is tangled in Equation (3), we calculate the higher derivative (third order here) of the position in Equation (7).

$$\begin{bmatrix} \dddot{x} \\ \dddot{y} \end{bmatrix} = \frac{1}{m} \cdot \dot{J}_{\psi+\alpha} \cdot J_\theta \cdot \begin{bmatrix} \omega_1^2 \\ \omega_2^2 \end{bmatrix} + \frac{1}{m} \cdot J_{\psi+\alpha} \cdot J_\theta \cdot \begin{bmatrix} 2\omega_1 \cdot \dot{\omega}_1 \\ 2\omega_2 \cdot \dot{\omega}_2 \end{bmatrix} \tag{7}$$

where $\dot{J}_{\psi+\alpha}$ is calculated in Equation (8).

$$\dot{J}_{\psi+\alpha} = \begin{bmatrix} -\sin(\psi+\alpha) & -\cos(\psi+\alpha) \\ \cos(\psi+\alpha) & -\sin(\psi+\alpha) \end{bmatrix} \cdot (\dot{\psi} + \dot{\alpha}) \tag{8}$$

Substituting Equation (2) into Equation (7) yields Equation (9).

$$\begin{bmatrix} \dddot{x} \\ \dddot{y} \end{bmatrix} = [\Delta_1^{2\times2} | \Delta_2^{2\times1}] \cdot \begin{bmatrix} \dot{\omega}_1 \\ \dot{\omega}_2 \\ \dot{\alpha} \end{bmatrix} \tag{9}$$

where $\Delta_1^{2\times2}$ and $\Delta_2^{2\times1}$ are defined in Equation (10)—(11).

$$\Delta_1^{2\times2} = \frac{1}{m} \cdot J_{\psi+\alpha} \cdot J_\theta \cdot \begin{bmatrix} 2\omega_1 & 0 \\ 0 & 2\omega_2 \end{bmatrix} \tag{10}$$

$$\Delta_2^{2\times1} = \frac{1}{m} \cdot \begin{bmatrix} -\sin(\psi+\alpha) & -\cos(\psi+\alpha) \\ \cos(\psi+\alpha) & -\sin(\psi+\alpha) \end{bmatrix} \cdot J_\theta \cdot \left(1 - \frac{I_T}{I_B}\right) \cdot \begin{bmatrix} \omega_1^2 \\ \omega_2^2 \end{bmatrix} \tag{11}$$

Since $\Delta_1^{2\times 2}$ is invertible if and only if $\omega_1 \neq 0$ and $\omega_2 \neq 0$, $[\Delta_1^{2\times 2}|\Delta_2^{2\times 1}]$ has full row rank if and only if $\omega_1 \neq 0$ and $\omega_2 \neq 0$. Thus, the pseudo-inverse of $[\Delta_1^{2\times 2}|\Delta_2^{2\times 1}]$ has explicit form, Equation (12).

$$pinv(\Delta) = \Delta^T \cdot (\Delta \cdot \Delta^T)^{-1} \quad (12)$$

where $\Delta = [\Delta_1^{2\times 2}|\Delta_2^{2\times 1}]$.

Separating the inputs in Equation (9), we receive Equation (13).

$$\begin{bmatrix} \dot{\omega}_1 \\ \dot{\omega}_2 \\ \dot{\alpha} \end{bmatrix} = pinv([\Delta_1^{2\times 2}|\Delta_2^{2\times 1}]) \cdot \begin{bmatrix} \dddot{x} \\ \dddot{y} \end{bmatrix} \quad (13)$$

Based on Equation (13), the controller is designed in Equation (14).

$$\begin{bmatrix} \dot{\omega}_1 \\ \dot{\omega}_2 \\ \dot{\alpha} \end{bmatrix} = pinv([\Delta_1^{2\times 2}|\Delta_2^{2\times 1}]) \cdot \begin{bmatrix} \dddot{x}_r + k_{x_1} \cdot (\ddot{x}_r - \ddot{x}) + k_{x_2} \cdot (\dot{x}_r - \dot{x}) + k_{x_3} \cdot (x_r - x) \\ \dddot{y}_r + k_{y_1} \cdot (\ddot{y}_r - \ddot{y}) + k_{y_2} \cdot (\dot{y}_r - \dot{y}) + k_{y_3} \cdot (y_r - y) \end{bmatrix} \quad (14)$$

where $x_r$ and $y_r$ are the position reference. $k_{x_i}$ and $k_{y_i}$ ($i = 1, 2, 3$) are the coefficients for the controller.

## 4. SIMULATION AND STATE DRIFT PHENOMENON

This section provides the simulation results of the control rules in Section 3. An interesting phenomenon, which we call state drift, in the result is paid attention to.

The reference is a circular trajectory starting from $(0, 0)$ in Figure 5. The radius of the reference is 10. The velocity of the reference is 10. The initial position of the vehicle is $(0, 0)$. The initial velocity, is also zero. The initial angular velocities are $\omega_1 = 200$ and $\omega_2 = 200$.

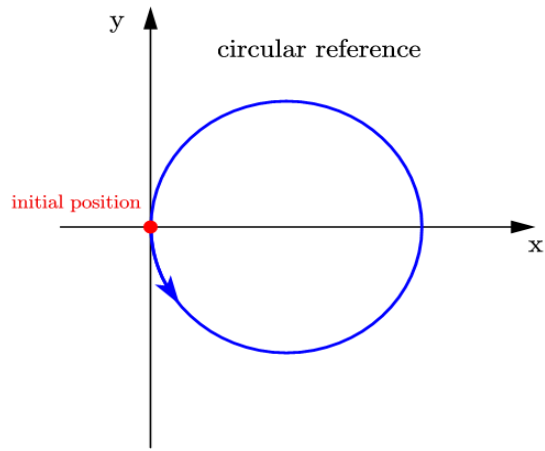

Fig. 5. The circular reference.

The solver is a fixed-step ODE3 in Simulink, MATLAB. The sampling time is 0.01 second.

## 4.1. Results for Third Derivative Feedback Linearization

The Simulink block diagram is plotted in Figure 6. The Dynamics part, Reference part, and Controller part are also shown.

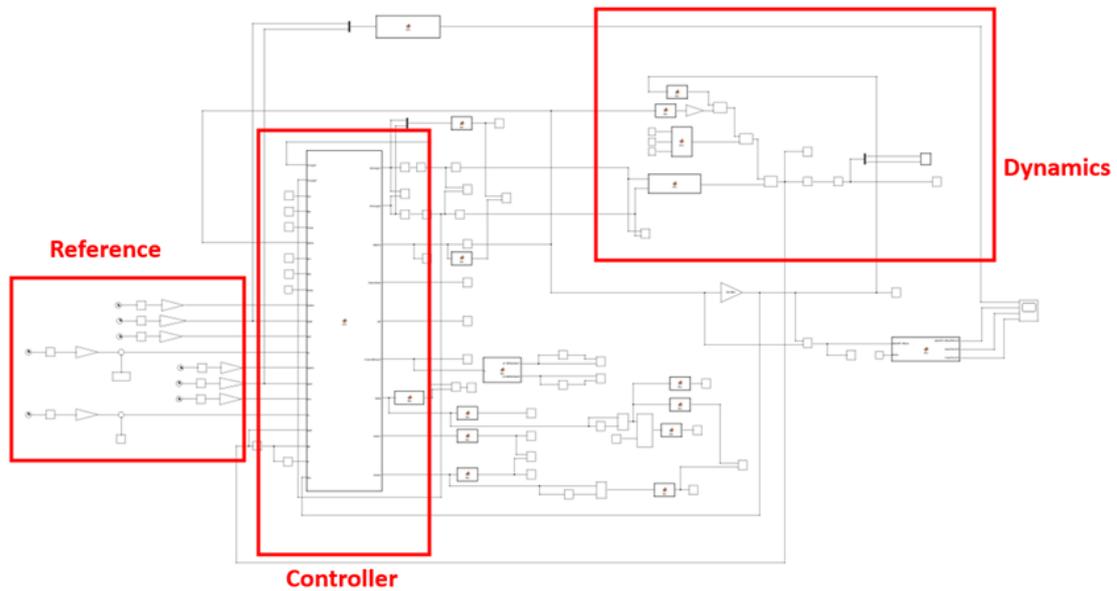

Fig. 6. Simulink block in simulation

Set the simulation time 10 seconds. The results are illustrated in Figure 7—10.

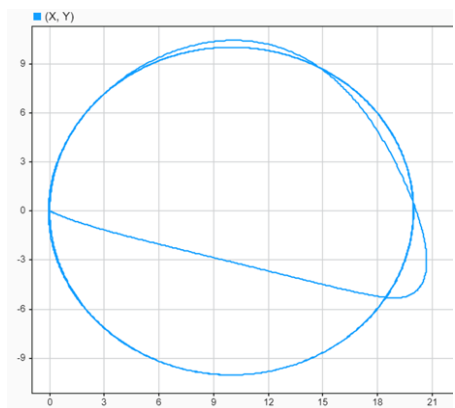 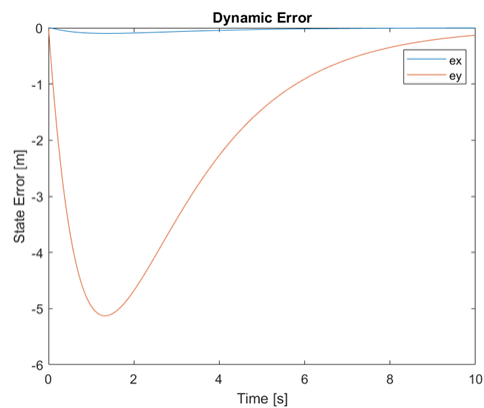

Fig. 7. The trajectory.		Fig. 8. The dynamic state error.

In Figure 7, the vehicle moves on a biased circular trajectory. The dynamic state error can be checked in Figure 8. Both the dynamic state errors in $x$ and $y$ directions decrease to zero.

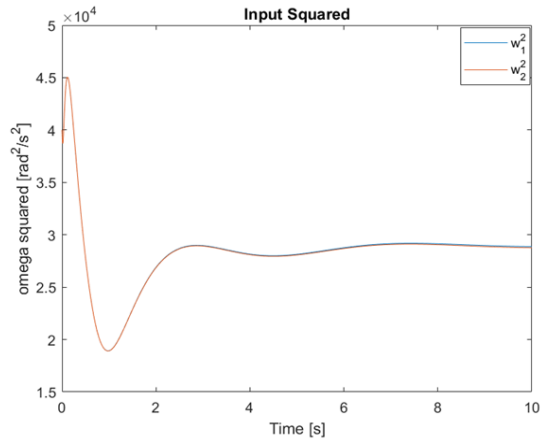

Fig. 9. Input squared.

Figure 9 shows the square of the input signals. Notice that two inputs almost overlap each other in Figure 9, especially at the beginning. The input signals seem to diverge/separate in the later time.

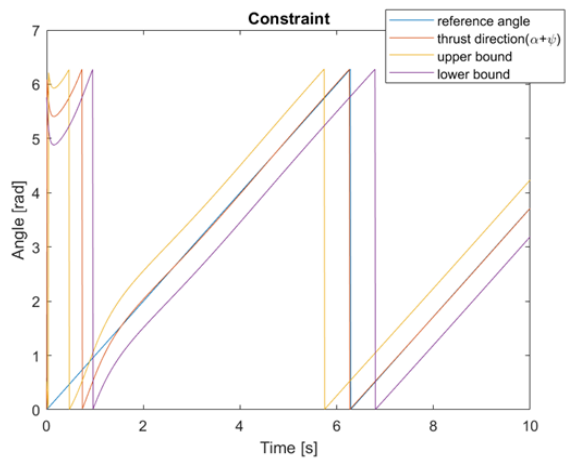

Fig. 10. The direction of the thrust.

Figure 10 plots the information related to the direction of the thrust. The direction of the middle of two thrusts can be calculated in Equation (15).

$$\alpha + \psi \quad (15)$$

The orange line in Figure 10 represents the direction of the middle of two thrusts. The yellow line (upper bound) and the purple line (lower bound) represent the admissible direction bound of the vehicle. The potential direction for generating the acceleration is between the lower bound and the upper bound. The upper bound and lower bound are computed in Equation (16)—(17), respectively.

$$\alpha + \psi + \frac{\pi}{6} \quad (16)$$

$$\alpha + \psi - \frac{\pi}{6} \qquad (17)$$

The blue line in Figure 10 is the direction pointing to the center of the circle (Figure 11). We can see that the direction of the middle of two thrusts of the vehicle (orange line) seems to stabilize at the direction pointing to the center of the circle (Figure 11) when time approaches 10 seconds.

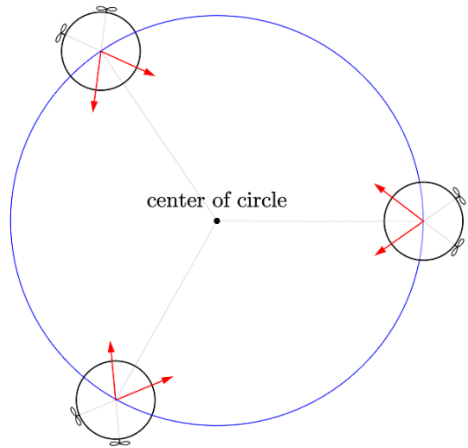

Fig. 11. The direction pointing to the center of the circle

Extending the simulation time to 2000 seconds, we receive the results in Figure 12—15.

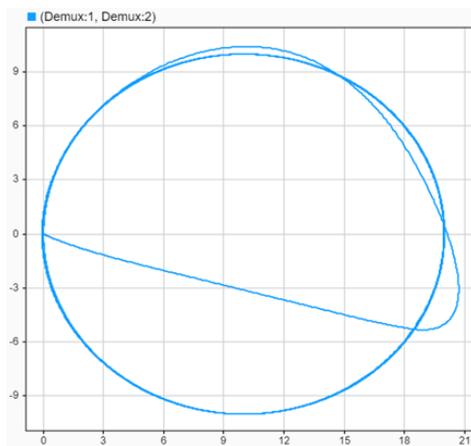 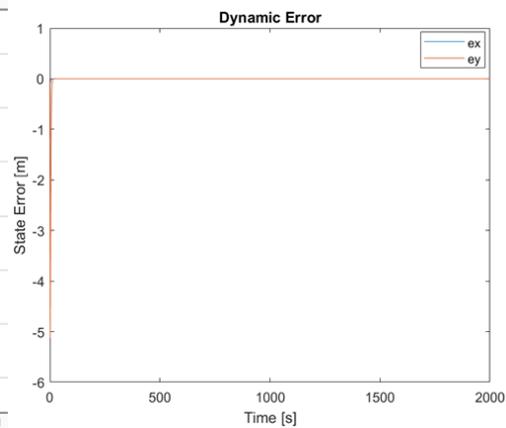

Fig. 12. The trajectory.          Fig. 13. The dynamic state error.

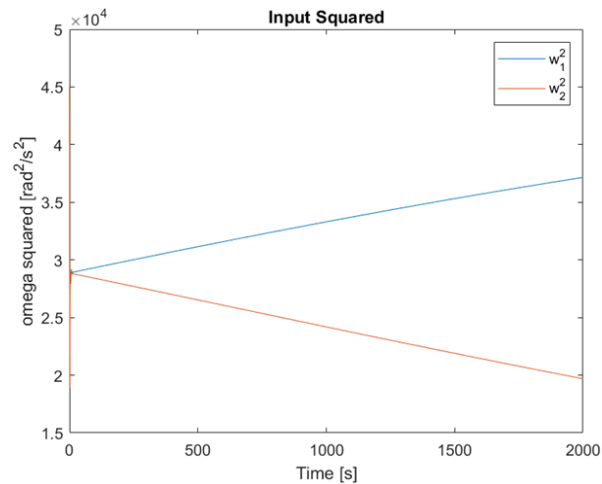

Fig. 14. Input squared.

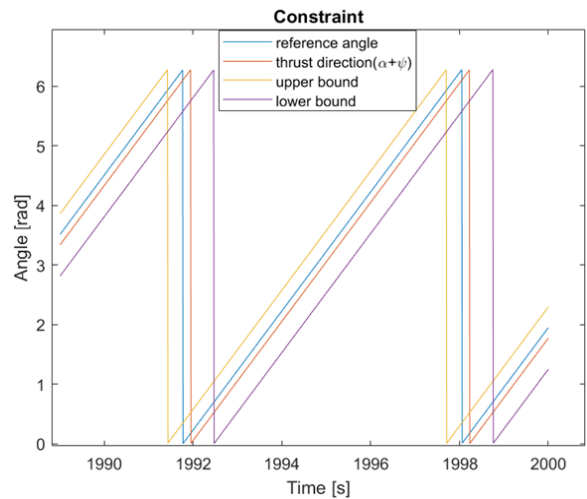

Fig. 15. The direction of the thrust.

The trajectory follows the circle in high precision (Figure 12,13) after being controlled for sufficient long time. The dynamic state error remains close to zero, indicating the stability of the system. Figure 14 shows the squares of the two angular velocities. The inputs diverge in two opposite directions.

Figure 15 shows the acceleration constraint around 1990—2000 seconds. The blue line in Figure 15 is the direction pointing to the center of the circle (Figure 11). The orange line is the direction of the middle of two thrusts. It dramatically biases from the direction pointing to the center of the circle (blue line).

### 4.2. State Drift

The interesting phenomenon in the result is that the input changes dramatically after the vehicle is stable already. We define this phenomenon 'State Drift'. State Drift happens only in the over-actuated systems. The direct reason causing State Drift refers to Equation (12), (14). While deducing it mathematically in a systematic way can be hard or even impossible.

While the underlying reason causing State Drift in this vehicle can be obviously explained.

The centripetal force pointing to the center of the circle is required to follow the reference in Figure 15. While generating a specific centripetal force, the combination of the magnitudes of the two thrusts $(F_1, F_2)$ is not unique (Figure 16—18). State Drift can be avoided if the number of the inputs is no more than the number of the output.

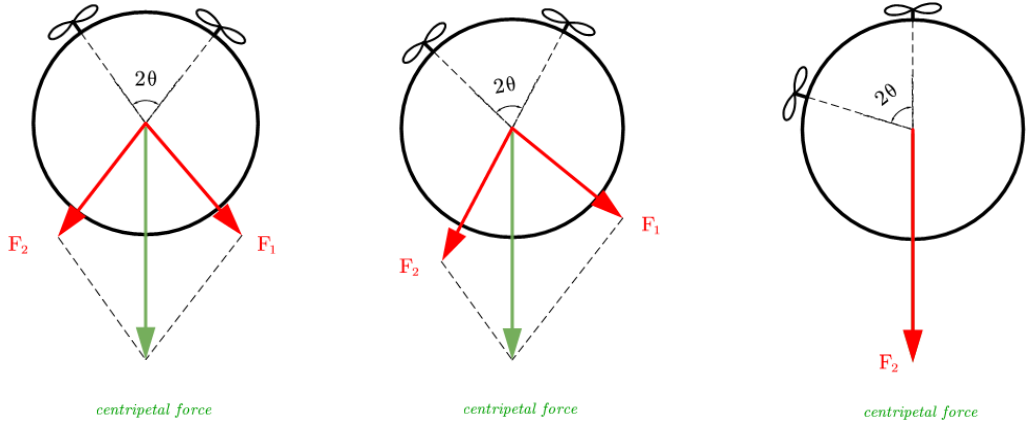

Fig. 16. Symmetrical.    Fig. 17. Symmetrical.    Fig. 18. Degrade to one thrust.

## 5. GAIT PLAN

The approach to avoid the State Drift is to decrease the number of inputs to be equal to the number of the outputs. In this research, we plan the direction of the middle of two thrusts in advance and discard the tilt angle, $\alpha$, from the input. The only two inputs then are the angular velocities of the two propellers, $\omega_1$ and $\omega_2$.

$\alpha$ is defined in Equation (18) such that the direction of the middle of two thrusts $(\psi + \alpha)$ always points to the center of the circular reference in Figure 11 during the entire movement.

$$\alpha = \frac{1}{1 - \frac{I_T}{T_B}} \cdot t \qquad (18)$$

Having decided $\alpha$ in Equation (18), only two inputs, $\omega_1$ and $\omega_2$, are left to be defined.

Based on the dynamics in Equation (3), the dynamic inversion is completed in Equation (19).

$$\begin{bmatrix} \omega_1^2 \\ \omega_2^2 \end{bmatrix} = m \cdot (J_{\psi+\alpha} \cdot J_\theta)^{-1} \cdot \begin{bmatrix} \ddot{x} \\ \ddot{y} \end{bmatrix} \qquad (19)$$

It is worth mentioning that Equation (19) can be received for the reason that the term $J_{\psi+\alpha} \cdot J_\theta$ is invertible.

Further, we develop the PD controllers in Equation (20).

$$\begin{bmatrix} \omega_1^2 \\ \omega_2^2 \end{bmatrix} = m \cdot (J_{\psi+\alpha} \cdot J_\theta)^{-1} \cdot \begin{bmatrix} \ddot{x}_r + k_{x_1} \cdot (\dot{x}_r - \dot{x}) + k_{x_2} \cdot (x_r - x) \\ \ddot{y}_r + k_{y_1} \cdot (\dot{y}_r - \dot{y}) + k_{y_2} \cdot (y_r - y) \end{bmatrix} \quad (20)$$

where $x_r$ and $y_r$ are the position reference. $k_{x_i}$ and $k_{y_i}$ $(i=1,2)$ are the coefficients for the controller.

The same reference in Figure 5 is used for the tracking experiment to test the controller in Equation (20). The Simulink block diagram for this experiment is plotted in Figure 19. The dynamics, controller, and the reference parts are marked.

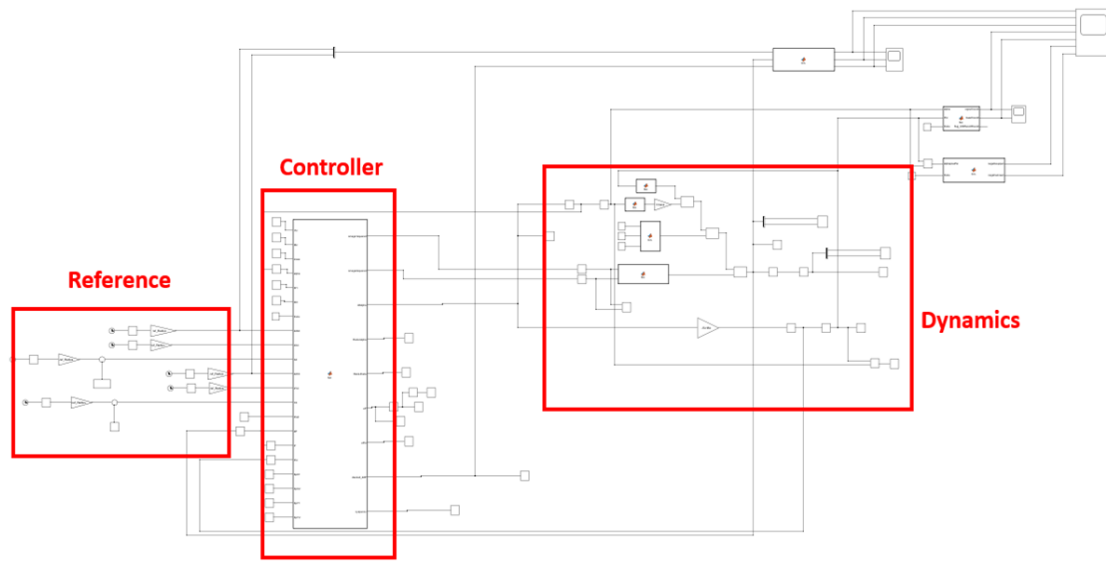

Fig. 19. Simulink block diagram

The simulation time is 10 seconds. The sampling time is 0.01 second (ODE3 fixed-step solver). The result is plotted in Figures 20—23.

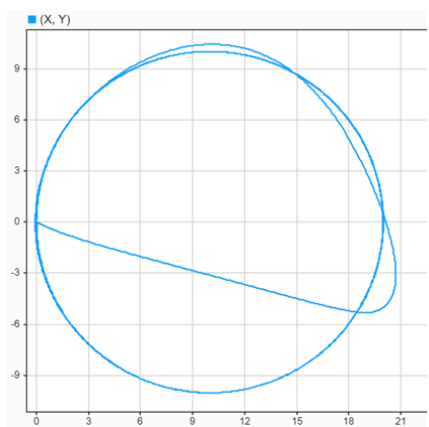
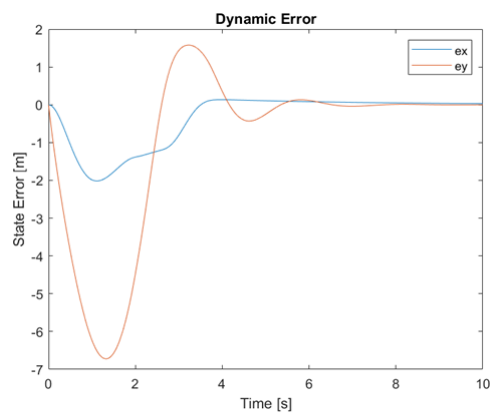

Fig. 20. The trajectory.  Fig. 21. The dynamic state error.

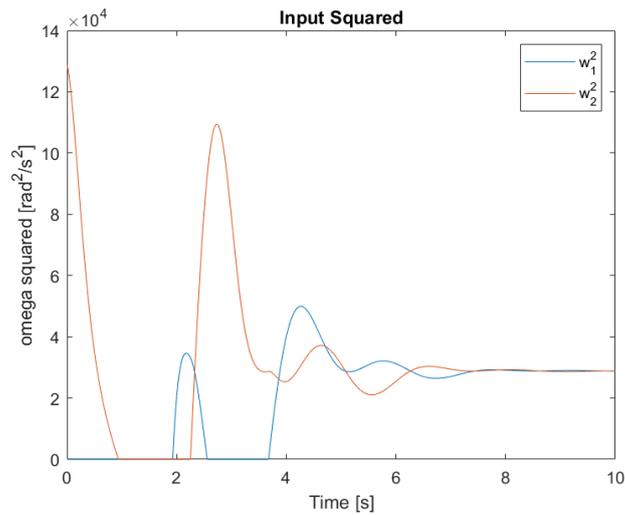

Fig. 22. Input squared.

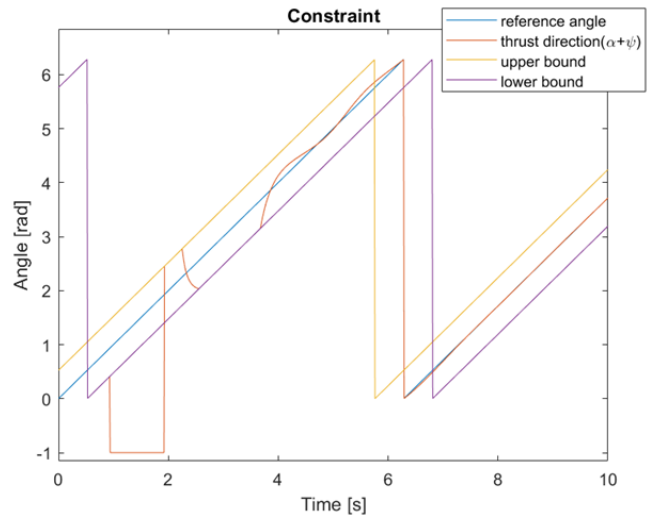

Fig. 23. The direction of the thrust (angle -1 represents 0 acceleration).

Judged from Figure 20—21, the trajectory generally follows the reference as a circle after some time. In Figure 23, we can see that the inputs do not diverge in two opposite directions. Actually, they become identical in the end. This is because that the number of the inputs is equal to the number of the outputs; State Drift can happen only in an actuated system. For a system with an equal number of inputs and outputs, State Drift will not happen.

There is an undesired behavior in this controller: the saturation in inputs.

The feedback linearization transfers the nonlinear system to a linear system (as what we have done in Equation (19)). So that the linear controller can be applied. This approach is valid and solid if and only if the constraint is not effective; hitting the bound (e.g., positive thrust constraint) invalidates the feedback linearization method.

The nonlinear property of the original system should be considered if the saturation happens. However, there lacks stability proof for a general linear time invariant system. Only several categories of linear time variant systems with specific forms can be proved stable [19,20]. Seeking the stability proof for the rest linear time variant systems is still an open question.

Unfortunately, when saturation happens in this controller, Equation (20), with the circular reference in Figure 5, the linear time variant system lies in the category whose stability proof is not found. The detail of the deducing process is omitted in this paper, since the analysis of a nonlinear system is beyond this research.

The saturation can also be observed in Figure 23, the direction of the thrust. It can be seen that the input signals saturate at around 3 seconds.

## 6. CONCLUSIONS AND DISCUSSIONS

The results in the over-actuated state feedback linearization show that the system is stabilized. The State Drift happens in the controlled over-actuated system, although the stability is not affected in this system.

Applying the gait plan to this system, the number of the inputs becomes equal to the number of the outputs. Subsequently, the State Drift phenomenon disappears. The saturation happens in this control method, which affects the stability proof.

Further steps can be deducing the stability proof for the control methods in gait plan with saturation and different gait patterns development.

In the experiments in this research, State Drift seems to only affect the allocation of the inputs. While the stability seems not be affected by this phenomenon. In this section, we will see a higher order feedback linearization controller. This controller is to stabilize the system in Equation (2)—(3). The State Drift in this example shows that the stability is affected.

### 6.1. Fourth Derivative Feedback Linearization

Calculate the fourth derivative of the position. It can be computed by differentiating Equation (7). The result yields:

$$\begin{bmatrix} \dddot{x} \\ \dddot{y} \end{bmatrix} = \frac{1}{m} \cdot J_\psi \cdot \ddot{J}_\alpha \cdot J_\theta \cdot \begin{bmatrix} \omega_1^2 \\ \omega_2^2 \end{bmatrix} + \frac{2}{m} \cdot J_\psi \cdot \dot{J}_\alpha \cdot J_\theta \cdot \begin{bmatrix} 2\omega_1 \cdot \dot\omega_1 \\ 2\omega_2 \cdot \dot\omega_2 \end{bmatrix} + \frac{1}{m} \cdot J_\psi \cdot J_\alpha \cdot J_\theta \cdot \begin{bmatrix} 2\dot\omega_1^2 + 2\omega_1 \cdot \ddot\omega_1 \\ 2\dot\omega_2^2 + 2\omega_2 \cdot \ddot\omega_2 \end{bmatrix} + \\ \frac{2}{m} \cdot \dot{J}_\psi \cdot \dot{J}_\alpha \cdot J_\theta \cdot \begin{bmatrix} \omega_1^2 \\ \omega_2^2 \end{bmatrix} + \frac{2}{m} \cdot \dot{J}_\psi \cdot J_\alpha \cdot J_\theta \cdot \begin{bmatrix} 2\omega_1 \cdot \dot\omega_1 \\ 2\omega_2 \cdot \dot\omega_2 \end{bmatrix} + \frac{1}{m} \cdot \ddot{J}_\psi \cdot J_\alpha \cdot J_\theta \cdot \begin{bmatrix} \omega_1^2 \\ \omega_2^2 \end{bmatrix}$$

(21)

where

$$J_\psi = \begin{bmatrix} \cos(\psi) & -\sin(\psi) \\ \sin(\psi) & \cos(\psi) \end{bmatrix}$$

(22)

$$J_\alpha = \begin{bmatrix} \cos(\alpha) & -\sin(\alpha) \\ \sin(\alpha) & \cos(\alpha) \end{bmatrix}$$

(23)

Equation (21) can also be written in Equation (24).

$$\begin{bmatrix} \dddot{x} \\ \dddot{y} \end{bmatrix} = M_P + \Delta_P \cdot \begin{bmatrix} \ddot{\omega}_1 \\ \ddot{\omega}_2 \\ \ddot{\alpha} \end{bmatrix} \tag{24}$$

where $\Delta_P \cdot \begin{bmatrix} \ddot{\omega}_1 \\ \ddot{\omega}_2 \\ \ddot{\alpha} \end{bmatrix}$ includes all the terms containing $\ddot{\omega}_1$, $\ddot{\omega}_2$, or $\ddot{\alpha}$. $M_P$ includes all the remaining terms without containing $\ddot{\omega}_1$, $\ddot{\omega}_2$, or $\ddot{\alpha}$.

Specifically,

$$\Delta_P = [\Delta_{P_\omega}^{2\times 2} | \Delta_{P_\alpha}^{2\times 1}] \tag{25}$$

Where

$$\Delta_{P_\omega}^{2\times 2} = \frac{1}{m} \cdot J_\psi \cdot J_\alpha \cdot J_\theta \cdot \begin{bmatrix} 2\omega_1 & 0 \\ 0 & 2\omega_2 \end{bmatrix} \tag{26}$$

$$\Delta_{P_\alpha}^{2\times 1} = \frac{1}{m} \cdot J_\psi \cdot \begin{bmatrix} -\sin(\alpha) & -\cos(\alpha) \\ \cos(\alpha) & -\sin(\alpha) \end{bmatrix} \cdot J_\theta \cdot \begin{bmatrix} \omega_1^2 \\ \omega_2^2 \end{bmatrix} \tag{27}$$

Notice that when $\omega_1$ and $\omega_2$ are not both zero, we have

$$Rank(\Delta_{P_\omega}^{2\times 2}) = 2 \tag{28}$$

Equation (28) indicates that $\Delta_P$ has full row rank when $\omega_1$ and $\omega_2$ are not both zero. With this condition, $pinv(\Delta_P)$ can be calculated based on Equation (12).

Separating the inputs in Equation (24) yields Equation (29).

$$\begin{bmatrix} \ddot{\omega}_1 \\ \ddot{\omega}_2 \\ \ddot{\alpha} \end{bmatrix} = pinv(\Delta_P) \cdot \left( \begin{bmatrix} \dddot{x} \\ \dddot{y} \end{bmatrix} - M_P \right) \tag{29}$$

PD controllers are built based on Equation (29).

The reference and the initial condition are identical to the settings in the previous experiments.

They are initial $\omega_1 = 200$, initial $\omega_2 = 200$, and Figure 5.

The dynamic state error and the input squared are in Figure 24 and 25, respectively.

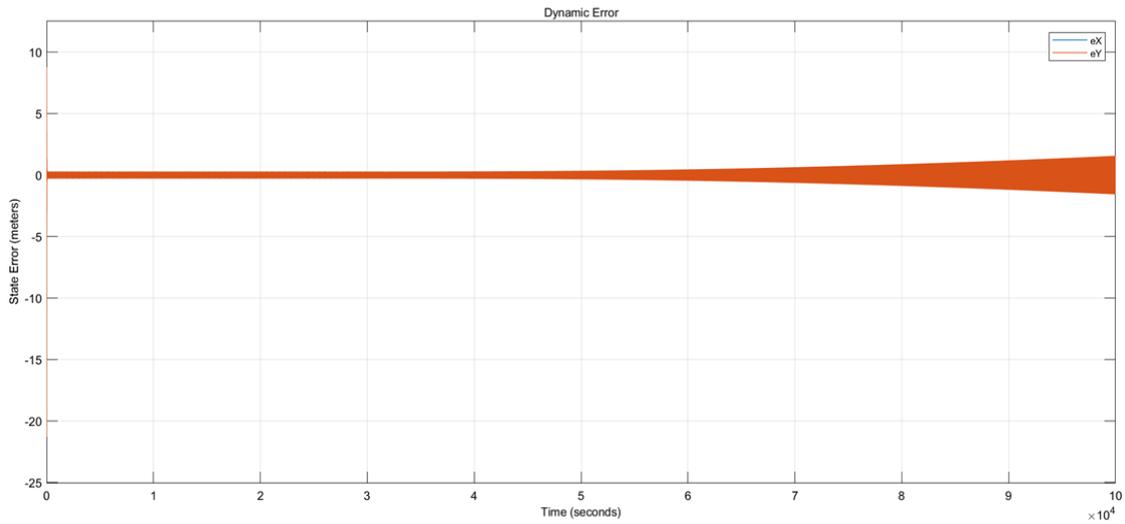

Fig. 24. Dynamic State Error.

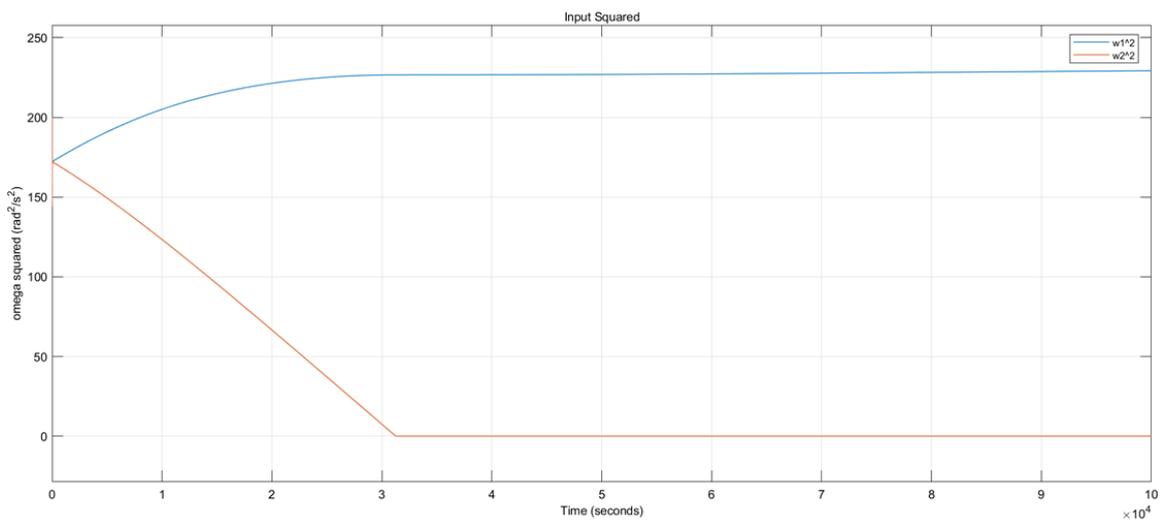

Fig. 25. Input Squared.

The dynamic state error (Figure 24) decreases to an acceptably small value (0.2 meter) at the beginning. While it increases after around $3 \times 10^4$ seconds. Finally, the dynamic state error diverges from 0, potentially making this system unstable.

State Drift causes this.

Judged from Figure 25, the State Drift happens so that the two inputs diverge in the two opposite directions. As we know from the previous experiments, it can not be a factor causing unstable.

However, after 3 seconds, the input ($\omega_2^2$) hit the positive constraint (negative thrust is not allowed), destroying the underlying stability proof. The consequence is what we see in Figure 24; unstable system with increasing dynamic state error is reported.

## 6.2. State Drift Phenomenon

So far, State Drift phenomenon has been thoroughly reported by controlling a fictional system (tilt vehicle). The reason can be multiple equilibrium states (inputs), which is explained in Figure 16—18.

However, the mathematical cause for pulling the system from one equilibrium to another is complicated and beyond the scope of this research.

For example, the following question is extremely hard to reply:

I understand that the system is stable even if State Drift happens, but why does State Drift happen?
The answer to this question can also be a challenged further step.

## AUTHORS

**Zhe Shen** received his M.S. degree from Technical University of Denmark, Denmark, in 2019. In 2020, he joined The University of Tokyo, where he is currently a Ph.D. student of the Department of Aeronautics and Astronautics.

**Takeshi Tsuchiya** received his M.E. and Ph.D. degrees from The University of Tokyo, Japan, in 1997 and 2000, respectively. In 2002, he joined The University of Tokyo, where he is currently a Professor of Department of Aeronautics and Astronautics.